
\def\ung{{{\frak{g}}}}
\def\ungh{{{{\hat{\frak{g}}}}}}
\def\uqg{{{U_{q}(\ung)}}}
\def\uqgh{{{U_{q}(\ungh)}}}

\def\bp{{{\bold{P}}}}
\def\bq{{{\bold{Q}}}}
\def\calp{{{{\Cal{P}}}}}
\def\calq{{{{\Cal{Q}}}}}
\def\ot{{{\otimes}}}
\def\op{{{\oplus}}}

\def\unh{{{\frak{h}}}}

\def\calp{\Cal P}

\def\vbp{v_{\bp}}

\def\ungsh{\ungh_J}
\def\uqgs{U_q(\ung_J)}
\def\uqgsh{U_q(\ungsh)}

\magnification 1200
\input amstex
\documentstyle{amsppt}
\document
\centerline{\bf{{ Minimal Affinizations of Representations}}}
\vskip 12pt
\centerline{\bf of Quantum Groups:}
\vskip 12pt
\centerline{\bf the Simply-laced Case}
\vskip 36pt
\centerline{Vyjayanthi Chari{\footnote{Both authors were partially supported by
the NSF, DMS--9207701}},}
\vskip 12pt
\centerline{Department of Mathematics,}
\vskip 12pt
\centerline{University of California, Riverside, CA 92521, USA.}
\vskip 24pt
\centerline{Andrew Pressley,}
\vskip 12pt
\centerline{Department of Mathematics,}
\vskip 12pt
\centerline{King's College, Strand, London WC2R 2LS, UK.}

\vskip 36pt
\noindent{\bf Introduction}
\vskip12pt\noindent In [2], we defined the notion of an
affinization of a finite-dimensional irreducible representation $V$ of the
quantum group $\uqg$, where $\ung$ is a finite-dimensional complex simple Lie
algebra and $q\in\Bbb C^\times$ is transcendental. An affinization of $V$ is an
irreducible representation $\hat{V}$ of the quantum affine algebra $\uqgh$
which, regarded as a representation of $\uqg$, contains $V$ with multiplicity
one, and is such that all other irreducible components of $\hat{V}$ are
strictly smaller than $V$, with respect to a certain natural partial order on
the set of isomorphism classes of finite-dimensional representations of $\uqg$.
In general, a given representation $V$ has finitely many affinizations up to
$\uqg$-isomorphism (always at least one), and it is natural to look for the
minimal one(s). We refer the reader to the introduction to [2] for a discussion
of the significance of the notion of an affinization.

In [2], we show that, if $\ung$ has rank 2, every $V$ has a unique minimal
affinization. In this paper, we consider the case when $\ung$ is a simply-laced
algebra of arbitrary rank. If $\ung$ is of type A, there is again a unique
minimal affinization (this result is, in fact, contained in [4]). But, if
$\ung$ is of type D or E, and if the highest weight of $V$ is not too singular,
we show that $V$ has precisely three minimal affinizations. In all cases, the
minimal affinization(s) are described precisely in terms of the parametrization
of the finite-dimensional irreducible representations of $\uqgh$ given in [3]
(in the $sl_2$ case), in [5] (in the $sl_n$ case), and in [6] (in the general
case).

\vskip36pt\noindent{\bf 1 Quantum affine algebras and their representations}
\vskip 12pt\noindent
In this section, we collect the results about quantum affine algebras which we
shall need later.

Let $\ung$ be a finite--dimensional complex simple Lie algebra with Cartan
subalgebra $\unh$ and Cartan matrix $A=(a_{ij})_{i,j\in I}$. Fix coprime
positive integers  $(d_i)_{i\in I}$\/ such that $(d_ia_{ij})$\/ is symmetric.
Let $P=\Bbb Z^I$ and let
$P^+=\{\lambda\in P\mid \lambda(i)\ge 0\ \text{for all $i\in I$}$\}. Let $R$
(resp. $R^+$) be the set of roots (resp. positive roots) of $\ung$. Let
$\alpha_i$ ($i\in I$) be the simple roots and let $\theta$ be the highest root.
Define a non-degenerate symmetric bilinear form $(\ ,\ )$ on $\unh^*$ by
$(\alpha_i,\alpha_j)=d_ia_{ij}$,
and set $d_0=\frac12(\theta,\theta)$. Let $Q = \op_{i\in I}\Bbb
Z.\alpha_i\subset\unh ^*$\/ be the root lattice, and set $Q^+ =\sum_{i\in
I}\Bbb N.\alpha_i$. Define a partial order $\ge$ on $P$ by $\lambda\ge \mu$ iff
$\lambda-\mu\in Q^+$.

Let $q\in \Bbb C^{\times}$ be transcendental, and, for $r,n\in\Bbb N$, $n\ge
r$, define
$$\align [n]_q & =\frac{q^n -q^{-n}}{q -q^{-1}},\\
[n]_q! &=[n]_q[n-1]_q\ldots [2]_q[1]_q,\\
\left[{n\atop r}\right]_q &= \frac{[n]_q!}{[r]_q![n-r]_q!}.\endalign$$

\proclaim{Proposition 1.1} There is a Hopf algebra $\uqg$ over $\Bbb C$ which
is generated as an algebra by elements $x_i^{{}\pm{}}$, $k_i^{{}\pm 1}$ ($i\in
I$), with the following defining relations:
$$\align
k_ik_i^{-1} = k_i^{-1}k_i &=1,\;\;  k_ik_j =k_jk_i,\\
k_ix_j^{{}\pm{}}k_i^{-1} &= q_i^{{}\pm a_{ij}}x_j^{{}\pm},\\
[x_i^+ , x_j^-] &= \delta_{ij}\frac{k_i - k_i^{-1}}{q_i -q_i^{-1}},\\
\sum_{r=0}^{1-a_{ij}}\left[{{1-a_{ij}}\atop r}\right]_{q_i}
(x_i^{{}\pm{}})^rx_j^{{}\pm{}}&(x_i^{{}\pm{}})^{1-a_{ij}-r} =0, \ \ \ \ i\ne
j.\endalign$$

The comultiplication $\Delta$, counit $\epsilon$, and antipode $S$ of $\uqg$
are given by
$$\align\Delta(x_i^+)&= x_i^+\ot k_i +1\ot x_i^+,\\
\Delta(x_i^-)&= x_i^-\ot 1 +k_i^{-1}\ot x_i^-,\\
\Delta(k_i^{{}\pm 1}) &= k_i^{{}\pm 1}\ot k_i^{{}\pm 1},\\
\epsilon(x_i^{{}\pm{}}) =0,\;\ & \epsilon(k_i^{{}\pm 1}) =1,\\
S(x_i^+) = -x_i^+k_i^{-1},\; S(x_i^-) &=- k_ix_i^-, \; S(k_i^{{}\pm 1})
=k_i^{{}\mp 1},\endalign$$
for all $i\in{I}$.\qed
\endproclaim

The Cartan involution $\omega$ of $\uqg$ is the unique algebra automorphism of
$\uqg$ which takes $x_i^{{}\pm{}}\mapsto -x_i^{{}\mp{}}$, $k_i^{{}\pm 1}\mapsto
k_i^{{}\mp 1}$, for all $i\in I$.

Let $\hat{I} = I\amalg\{0\}$\/ and let ${\hat A} =(a_{ij})_{i,j\in {\hat I}}$
be the extended Cartan matrix of $\ung$, i.e. the generalized Cartan matrix of
the (untwisted) affine Lie algebra $\hat\ung$ associated to $\ung$. Let
$q_0=q^{d_0}$.

\proclaim{Theorem 1.2} Let $\uqgh$ be the algebra with generators
$x_i^{{}\pm{}}$, $k_i^{{}\pm 1}$ ($i\in\hat{I}$) and defining relations those
in 1.1, but with the indices $i$, $j$ allowed to be arbitrary elements of $\hat
I$. Then, $\uqgh$ is a Hopf algebra with comultiplication, counit and antipode
given by the same formulas as in 1.1 (but with $i\in\hat{I}$).

Moreover, $\uqgh$ is isomorphic to the algebra ${\Cal A}_q$ with generators
$x_{i,r}^{{}\pm{}}$ ($i\in I$, $r\in\Bbb Z$), $k_i^{{}\pm 1}$ ($i\in I$),
$h_{i,r}$ ($i\in I$, $r\in \Bbb Z\backslash\{0\}$) and $c^{{}\pm{1/2}}$, and
the following defining relations:
$$\align
c^{{}\pm{1/2}}\ &\text{are central,}\\
k_ik_i^{-1} = k_i^{-1}k_i =1,\;\; &c^{1/2}c^{-1/2} =c^{-1/2}c^{1/2} =1,\\
k_ik_j =k_jk_i,\;\; &k_ih_{j,r} =h_{j,r}k_i,\\
k_ix_{j,r}k_i^{-1} &= q_i^{{}\pm a_{ij}}x_{j,r}^{{}\pm{}},\\
[h_{i,r} , x_{j,s}^{{}\pm{}}] &= \pm\frac1r[ra_{ij}]_{q_i}c^{{}\mp
{|r|/2}}x_{j,r+s}^{{}\pm{}},\\
x_{i,r+1}^{{}\pm{}}x_{j,s}^{{}\pm{}} -q_i^{{}\pm
a_{ij}}x_{j,s}^{{}\pm{}}x_{i,r+1}^{{}\pm{}} &=q_i^{{}\pm
a_{ij}}x_{i,r}^{{}\pm{}}x_{j,s+1}^{{}\pm{}}
-x_{j,s+1}^{{}\pm{}}x_{i,r}^{{}\pm{}},\\
[x_{i,r}^+ , x_{j,s}^-]=\delta_{ij} & \frac{ c^{(r-s)/2}\phi_{i,r+s}^+ -
c^{-(r-s)/2} \phi_{i,r+s}^-}{q_i - q_i^{-1}},\\
\sum_{\pi\in\Sigma_m}\sum_{k=0}^m(-1)^k\left[{m\atop k}\right]_{q_i} x_{i,
r_{\pi(1)}}^{{}\pm{}}\ldots x_{i,r_{\pi(k)}}^{{}\pm{}} & x_{j,s}^{{}\pm{}}
 x_{i, r_{\pi(k+1)}}^{{}\pm{}}\ldots x_{i,r_{\pi(m)}}^{{}\pm{}} =0,\ \ i\ne j,
\endalign$$
for all sequences of integers $r_1,\ldots, r_m$, where $m =1-a_{ij}$,
$\Sigma_m$ is the symmetric group on $m$ letters, and the
$\phi_{i,r}^{{}\pm{}}$ are determined by equating powers of $u$ in the formal
power series
$$\sum_{r=0}^{\infty}\phi_{i,\pm r}^{{}\pm{}}u^{{}\pm r} = k_i^{{}\pm 1}
exp\left(\pm(q_i-q_i^{-1})\sum_{s=1}^{\infty}h_{i,\pm s} u^{{}\pm s}\right).$$

If $\theta =\sum_{i\in I}m_i\alpha_i$, set $k_{\theta} = \prod_{i\in
I}k_i^{m_i}$. Suppose that the root vector $\overline{x}_{\theta}^+$ of $\ung$
corresponding to $\theta$ is expressed in terms of the simple root vectors
$\overline{x}_i^+$ ($i\in I$) of $\ung$ as
$$\overline{x}_{\theta}^+ = \lambda[\overline{x}_{i_1}^+, [\overline
x_{i_2}^+,\cdots ,[\overline x_{i_k}^+, \overline x_j^+]\cdots ]]$$
for some $\lambda\in\Bbb C^{\times}$. Define maps $w_i^{{}\pm{}}:\uqgh\to\uqgh$
by
$$w_i^{{}\pm{}}(a) = x_{i,0}^{{}\pm{}}a - k_i^{{}\pm 1}ak_i^{{}\mp
1}x_{i,0}^{{}\pm{}}.$$
Then, the isomorphism $f:\uqgh\to\Cal A_q$ is defined on generators by
$$\align
f(k_0) = k_{\theta}^{-1}, \ f(k_i) &= k_i, \ f(x_i^{{}\pm{}}) =
x_{i,0}^{{}\pm{}},  \ \ \ \ (i\in I),\\
f(x_0^+) &=\mu w_{i_1}^-\cdots w_{i_k}^-(x_{j,1}^-)k_{\theta}^{-1},\\
f(x_0^-) &=\lambda k_{\theta} w_{i_1}^+\cdots w_{i_k}^+(x_{j,-1}^+),\endalign
$$
where $\mu\in\Bbb C^{\times}$ is determined by the condition
$$[x_0^+, x_0^-] =\frac{k_0-k_0^{-1}}{q_0-q_0^{-1}}. \qed$$
\endproclaim

See [1], [5] and [7] for further details.

Note that there is a canonical homomorphism $\uqg\to\uqgh$ such that
$x_i^{{}\pm{}}\mapsto x_i^{{}\pm{}}$, $k_i^{{}\pm 1}\mapsto k_i^{{}\pm 1}$ for
all $i\in I$. Thus, any representation of $\uqgh$ may be regarded as a
representation of $\uqg$.

Let $\hat U^{{}\pm{}}$ (resp. $\hat U^0$) be the subalgebra of  $\uqgh$
generated by the $x_{i,r}^{{}\pm{}}$ (resp. by the $\phi_{i,r}^{{}\pm{}}$) for
all $i\in I$, $r\in\Bbb Z$. Similarly, let $U^{{}\pm{}}$ (resp. $U^0$) be the
subalgebra of $\uqg$ generated by the $x_i^{{}\pm{}}$ (resp. by the $k_i^{{}\pm
1}$) for all $i\in I$.
\proclaim{ Proposition 1.3} (a) $\uqg = U^-.U^0.U^+.$

(b) $\uqgh = \hat U^-.\hat U^0.\hat U^+.$ \qed
\endproclaim
See [5] or [8] for details.

We shall make use of the following automorphisms of $\uqgh$:
\proclaim{Proposition 1.4} (a) For all $t\in\Bbb C^\times$, there exists a Hopf
algebra automorphism $\tau_t$ of $\uqgh$ such that
$$\align
\tau_t(x_{i,r}^{{}\pm{}})&=t^rx_{i,r}^{{}\pm{}},\ \
\tau_t(h_{i,r})=t^rh_{i,r},\\
\tau_t(k_i^{{}\pm{1}})&=k_i^{{}\pm{1}},\ \
\tau_t(c^{{}\pm{1/2}})=c^{{}\pm{1/2}}.
\endalign$$
(b) There is a unique algebra involution $\hat\omega$ of $\uqgh$ given on
generators by
$$\aligned
\hat\omega(x_{i,r}^{{}\pm{}})=-x_{i,-r}^{{}\mp{}},\ \
&\hat\omega(h_{i,r})=-h_{i,r},\\
\hat\omega(\phi_{i,r}^{{}\pm{}})=\phi_{i,-r}^{{}\mp{}},\ \
&\hat\omega(k_i^{{}\pm 1})=k_i^{{}\mp 1},\\
\hat\omega(c^{{}\pm{1/2}})&=c^{{}\mp{1/2}}.
\endaligned$$
Moreover, we have
$$(\hat\omega\ot\hat\omega)\circ\Delta=\Delta^{op}\circ\hat\omega,$$
where $\Delta^{op}$ is the opposite comultiplication of $\uqgh$. \qed
\endproclaim
See [2] for the proof. Note that $\hat\omega$ is compatible, via the canonical
homomorphism $\uqg\to\uqgh$, with the Cartan involution $\omega$ of $\uqg$.

A representation $W$ of $\uqg$ is said to be of type 1 if it is the direct sum
of its weight spaces
$$W_{\lambda} =\{w\in W\mid k_i.w = q_i^{\lambda(i)}w\},\ \ \ \ \ (\lambda\in
P).$$
If $W_\lambda\ne 0$, then $\lambda$ is a weight of $W$. A vector $w\in
W_{\lambda}$ is a highest weight vector if $x_i^+.w =0$ for all $i\in I$, and
$W$ is a highest weight representation with highest weight $\lambda$ if
$W=\uqg.w$ for some highest weight vector $w\in W_\lambda$. Lowest weight
vectors and representations are defined similarly, by replacing $x_i^+$ by
$x_i^-$.

For a proof of the following proposition, see [5] or [8].
\proclaim{Proposition 1.5} (a) Every finite--dimensional representation of
$\uqg$ is completely reducible.

(b) Every finite--dimensional irreducible representation of $\uqg$ can be
obtained from a type 1 representation by twisting with an automorphism of
$\uqg$.

(c) Every finite--dimensional irreducible representation of $\uqg$ of type 1 is
both highest and lowest weight. Assigning to such a representation its highest
weight defines a bijection between the set of isomorphism classes of
finite--dimensional irreducible type 1 representations of $\uqg$ and $P^+$.

(d) The finite--dimensional irreducible representation $V(\lambda)$ of $\uqg$
of highest weight $\lambda\in P^+$ has the same character as the irreducible
representation of $\ung$ of the same highest weight.

(e) The multiplicity $m_\nu(V(\lambda)\ot V(\mu))$ of $V(\nu)$
in the tensor product $V(\lambda)\ot V(\mu)$, where $\lambda ,\mu, \nu\in P^+$,
is the same as in the tensor product of the irreducible representations of
$\ung$ of the same highest weight (this statement makes sense in view of parts
(a) and (c)).
\qed\endproclaim

A representation $V$ of $\uqgh$ is of type 1 if $c^{1/2}$ acts as the identity
on $V$, and if $V$ is of type 1 as a representation of $\uqg$. A vector $v\in
V$ is a highest weight vector if
$$x_{i,r}^+.v=0,\ \ \phi_{i,r}^{{}\pm{}}.v=\Phi_{i,r}^{{}\pm{}}v,\ \ \ c^{1/2}.
v =v,$$
for some complex numbers $\Phi_{i,r}^{{}\pm{}}$. A type 1 representation $V$ is
a highest weight representation if $V=\uqgh.v$, for some highest weight vector
$v$, and the pair of $(I\times\Bbb Z)$-tuples $(\Phi_{i,r}^{{}\pm{}})_{i\in
I,r\in\Bbb Z}$ is its highest weight. Note that $\Phi_{i,r}^+=0$ (resp.
$\Phi_{i,r}^-=0$) if $r<0$ (resp. if $r>0$), and that
$\Phi_{i,0}^+\Phi_{i,0}^-=1$. (In [5], highest weight representations of
$\uqgh$ are called `pseudo-highest weight'.) Lowest weight vectors and
representations of $\uqgh$ are defined similarly.

If $\lambda\in P^+$, let ${\Cal P}^\lambda$ be the set of all $I$-tuples
$(P_i)_{i\in I}$ of polynomials $P_i\in\Bbb C[u]$, with constant term 1, such
that $deg(P_i)=\lambda(i)$ for all $i\in I$. Set ${\Cal P}=\cup_{\lambda\in
P^+}{\Cal P}^\lambda$.

\proclaim{Theorem 1.6} (a) Every finite-dimensional irreducible representation
of $\uqgh$ can be obtained from a type 1 representation by twisting with an
automorphism of $\uqgh$.

(b) Every finite-dimensional irreducible representation of $\uqgh$ of type 1 is
both highest and lowest weight.

(c) Let $V$ be a finite-dimensional irreducible representation of $\uqgh$ of
type 1 and highest weight $(\Phi_{i,r}^{{}\pm{}})_{i\in I,r\in\Bbb Z}$. Then,
there exists $\bp=(P_i)_{i\in I}\in\calp$ such that
$$\sum_{r=0}^\infty\Phi_{i,r}^+u^r=q_i^{deg(P_i)}\frac{P_i(q_i^{-2}u)}{P_i(u)}=\sum_{r=0}^\infty\Phi_{i,r}^-u^{-r},$$
in the sense that the left- and right-hand terms are the Laurent expansions of
the middle term about $0$ and $\infty$, respectively. Assigning to $V$ the
$I$-tuple $\bp$ defines a bijection between the set of isomorphism classes of
finite-dimensional irreducible representations of $\uqgh$ of type 1 and
$\calp$.

(d) Let $\bp$, $\bq\in\calp$ be as above, and let $v_{\bp}$ and $v_\bq$ be
highest weight vectors of $V(\bp)$ and $V(\bq)$, respectively. Then, in
$V(\bp)\ot V(\bq)$,
$$x_{i,r}^+.(v_\bp\ot v_\bq)=0,\ \ \phi_{i,r}^{{}\pm{}}.(v_\bp\ot
v_{\bq})=\Psi_{i,r}^{{}\pm{}}(v_\bp\ot v_\bq),$$
where the complex numbers $\Psi_{i,r}^{{}\pm{}}$ are related to the polynomials
$P_iQ_i$ as the $\Phi_{i,r}^{{}\pm{}}$ are related to $P_i$ in (5). In
particular, if $\bp\ot\bq$ denotes the $I$-tuple $(P_iQ_i)_{i\in I}$, then
$V(\bp\ot\bq)$ is isomorphic to a quotient of the subrepresentation of
$V(\bp)\ot V(\bq)$ generated by the tensor product of the highest weight
vectors. \qed\endproclaim

See [5] for further details. If the highest weight
$(\Phi_{i,r}^{{}\pm{}})_{i\in I,r\in\Bbb Z}$ of $V$ is given by an $I$-tuple
$\bp$ as in part (c), we shall often abuse notation by saying that $V$ has
highest weight $\bp$.

We shall need the following result from [2].

\proclaim{Lemma 1.7} Let $\rho:\uqgh\to End(V)$ be a finite-dimensional
irreducible representation of type 1 with highest weight $\bp=(P_i)_{i\in I}$.
For any $t\in\Bbb C^\times$, denote by $\tau_t^*(V)$ the representation
$\rho\circ\tau_t$. Then, $\tau_t^*(V)$ has highest weight $\bp^t=(P_i^t)_{i\in
I}$, where
$$P_i^t(u)=P_i(tu).\qed$$
\endproclaim

Following [2], we say that a finite-dimensional irreducible representation $V$
of $\uqgh$ is an affinization of $\lambda\in P^+$ if $V\cong V(\bp)$ as a
representation of $\uqgh$, for some $\bp\in{\Cal P}^\lambda$. Two affinizations
of $\lambda$ are equivalent if they are isomorphic as representations of
$\uqg$; we denote by $[V]$ the equivalence class of $V$. Let ${\Cal Q}^\lambda$
be the set of equivalence classes of affinizations of $\lambda$.

The following result is proved in [2].
\proclaim{Proposition 1.8} If $\lambda\in P^+$ and $[V]$,
$[W]\in\Cal{Q}^{\lambda}$, we write $[V]\preceq [W]$ iff, for all $\mu\in P^+$,
either,

(i) $m_{\mu}(V)\le m_{\mu}(W)$, or

(ii) there exists $\nu>\mu$ with $m_{\nu}(V) < m_{\nu}(W)$.

\noindent Then, ${}\preceq{}$ is a partial order on ${\Cal
Q}^\lambda$.\qed\endproclaim

An affinization $V$ of $\lambda$ is minimal if $[V]$ is a minimal element of
$\Cal{Q}^{\lambda}$ for the partial order $\preceq$, i.e. if
$[W]\in\Cal{Q}^{\lambda}$ and $[W]\preceq [V]$ implies that $[V]=[W]$.
It is proved in [2] that ${\Cal Q}^\lambda$ is a finite set, so minimal
affinizations certainly exist.

\vskip 36pt\noindent
{\bf 2 Diagram subalgebras}
\vskip12pt\noindent
In this section, $\ung$ is any finite--dimensional complex simple Lie algebra.

Let $J$ be any non--empty connected subset of $I$, and let $\uqgs$ be the Hopf
subalgebra of $\uqg$ defined by the generators and relations in 1.1 for which
all the indices $i,j\in J$. Similarly, let $\uqgsh$ be the subalgebra of
$\uqgh$ defined by the generators and relations in 1.2 for which all the
indices $i,j\in J$.  Let $P_J$ be the set of weights of $\uqgs$, $R_J^+$ the
set of positive roots, etc. If $\lambda\in P$, let $\lambda_J$ be the
restriction of $\lambda: I\to \Bbb Z$ to $J$. Similarly, if $\bp =(P_i)_{i\in
I}\in\Cal P$ is an $I$--tuple of
polynomials in $\Bbb C[u]$ with constant term 1, let $\bp_J\in\Cal{P}_J$ be the
$J$--tuple $(P_i)_{i\in J}$.

Let $\Delta_J$ be the comultiplication of $\uqgsh$. Note that $\uqgsh$ is {\it
not} a Hopf subalgebra of $\uqgh$ in general. However, we do have
\proclaim{Lemma 2.1} Let $\emptyset\ne J\subseteq I$ be connected, and let
$\rho_J:\uqgsh\to\uqgh$ be the canonical homomorphism of algebras.
Then, for all $i\in J$,
$$\Delta(x_{i,{{}\pm 1}}^{{}\pm{}}) -(\rho_J\ot\rho_J)(\Delta_{J}(x_{i,{{}\pm
1}}^{{}\pm{}}))\in\bigoplus_{\eta',\eta''}\uqgh_{\eta'}\ot\uqgh_{\eta''},$$
where the sum is over those $\eta',\eta''\in Q\backslash Q_J$ such that $\eta'
+\eta'' ={{}\pm \alpha_i}$, and
$$\uqgh_{\eta}=\{u\in\uqgh|k_juk_j^{-1} =q^{\eta(j)}u \ \text{for all $j\in
I$}\}.\qed$$
\endproclaim

The proof of this lemma can be deduced in a straightforward manner from [1].

Fix a non-empty connected subset $J\subseteq I$. Let $\lambda\in P^+$,
$\bp\in\calp^\lambda$, and let $M$ be a highest weight representation of
$\uqgh$ with highest weight $\bp$ and highest weight vector $m$. Let
$M_J=U_q(\hat{\ung}_J).m$. Then, it follows from 1.3 that
$$M_J=\bigoplus_{\eta\in Q_J^+}M_{\lambda-\eta}.\tag1$$
Similarly, let $\mu\in P^+$, $\bq\in\calp^\mu$, let $N$ be a highest weight
representation of $\uqgh$ of highest weight $\bq$ and highest weight vector
$n$, and let $N_J=U_q(\hat{\ung}_J).n$. Then, we have
$$M_J\ot N_J=\bigoplus_{\eta\in Q_J^+}(M\ot N)_{\lambda+\mu-\eta}.\tag2$$
Indeed, it is obvious that the left-hand side of (2) is contained in the
right-hand side. On the other hand,
$$(M\ot N)_{\lambda+\mu-\eta}=\bigoplus_{\eta',\eta''}M_{\lambda-\eta'}\ot
N_{\mu-\eta''},$$
where the sum is over those $\eta'$, $\eta''\in Q^+$ such that
$\eta'+\eta''=\eta$. But, since $\eta\in Q_J^+$, this clearly forces $\eta'$,
$\eta''\in Q_J^+$, so by (1), $(M\ot N)_{\lambda+\mu-\eta}\subseteq M_J\ot
N_J$. This proves (2).

Now, $M_J\ot N_J$ admits an obvious action of $U_q(\hat{\ung}_J)$ by using
$\Delta_J$; we denote this representation by $M_J\ot_J N_J$. On the other hand,
for weight reasons, the action of the $\Delta(x_{i,r}^{{}\pm{}})$,
$\Delta(\phi_{i,r}^{{}\pm{}})$, for all $i\in J$, $r\in\Bbb Z$, obviously
preserves $\oplus_{\eta\in Q_J^+}(M\ot N)_{\lambda+\mu-\eta}$. This gives
another representation of  $U_q(\hat{\ung}_J)$ on $M_J\ot N_J$, using $\Delta$,
which we denote by $M_J\ot N_J$.

\proclaim{Proposition 2.2} The identity map $M_J\ot_J N_J\to M_J\ot N_J$ is an
isomorphism of representations of $U_q(\hat{\ung}_J)$.\endproclaim
\demo{Proof} The map obviously commutes with the action of $U_q({\ung}_J)$.
{}From 1.2, it follows that $U_q(\hat{\ung}_J)$ is generated as an algebra by
the elements of $U_q({\ung}_J)$, the $x_{i,r}^{{}\pm{}}$ for $i\in J$,
$r={{}\pm 1}$, and the $c^{{}\pm{1/2}}$. Since $c^{1/2}$ acts as the identity
on $M$ and $N$, it suffices to prove that, for all $m'\in M_J$, $n'\in N_J$,
$i\in J$, $r={{}\pm 1}$,
$$\Delta(x_{i,r}^{{}\pm{}}).(m'\ot n')
-(\rho_J\ot\rho_J)(\Delta_J(x_{i,r}^{{}\pm{}})).(m'\ot n')=0.\tag3$$
The left-hand side of (3) obviously belongs to $M_J\ot N_J$, since both terms
involved do. On the other hand, by 2.1, the left-hand side also belongs to
$$\bigoplus_{\eta',\eta''}\uqgh_{\eta'}.m'\ot\uqgh_{\eta''}.n',$$
where the sum is over those $\eta'$, $\eta''\in Q\backslash Q_J$ such that
$\eta'+\eta''={{}\pm{\alpha_i}}$. We may assume that $m'\in M_{\lambda-\xi'}$,
$n'\in N_{\mu-\xi''}$, where $\xi'$, $\xi''\in Q_J^+$. Then, the weight of the
first factor in a typical non-zero term in the above sum is
$\lambda-\xi'+\eta'$. On the other hand, by (1), its weight must be of the form
$\lambda-\eta$ for some $\eta\in Q_J^+$. Thus,
$$\eta'=\xi'-\eta.$$
But this is impossible, since $\xi'-\eta\in Q_J^+$ but $\eta'\notin Q_J^+$.
Hence, the left-hand side of (3) is zero.\qed\enddemo

\proclaim {Lemma 2.3} Let $\emptyset\ne J\subseteq I$ define a connected
subdiagram of the Dynkin diagram of $\ung$ .
Let $\bp\in\calp$, and let $v_{\bp}$ be a $\uqgh$--highest weight vector in
$V(\bp)$. Then, $\uqgsh.v_{\bp}$ is an irreducible  representation of $\uqgsh$
with highest weight $\bp_J$. \endproclaim
\demo{Proof} Let $W$ be a non--zero irreducible $\uqgsh$--subrepresentation of
$\uqgsh.v_{\bp}$. Since $\uqgsh.v_{\bp}$ is obviously preserved by the action
of $k_i$ for all $i\in I$, it follows  by 1.3 and 1.6(b) that we can choose
$0\ne w\in W\cap V(\bp)_{\mu}$, for some $\mu\in\lambda-Q_J^+$, such that
$$\align x_{i,r}^+. w &=0,\tag4\\
\phi_{i,r}^{{}\pm{}}.w &= \Phi_{i,r}^{{}\pm{}} w,\tag5\endalign$$
for some $\Phi_{i,r}^{{}\pm{}}\in\Bbb C$ and all $i\in J$, $r\in\Bbb Z$.
Since $\mu\in\lambda-Q_J^+$, we see that (1) actually holds for all $i\in I$,
$r\in\Bbb Z$. Let $W^+$ be the linear subspace spanned by all elements
$w\in\uqgsh.v_{\bp}\cap V(\bp)_{\mu}$ satisfying (4) and (5) for fixed
$\Phi_{i,r}^{{}\pm{}}$. The relations in 1.2 show that the
$\phi_{i,r}^{{}\pm{}}$ preserve $W^+$ for all $i\in I$, $r\in\Bbb Z$. Since the
$\phi_{i,r}^{{}\pm{}}$ act as commuting operators on $V(\bp)$, and so on $W^+$,
there exists $w'\in W^+$ satisfying both (4) and (5) for all $i\in I$, $r\in
\Bbb Z$. This means that $w'$ must be a scalar multiple of $v_{\bp}$, and so
$\mu=\lambda$. Thus, $W^+ =\Bbb C.v_{\bp}$ and the lemma is established.
\qed\enddemo

\proclaim{Lemma 2.4} Let $\emptyset\ne J\subseteq I$ define a connected
subdiagram of the Dynkin diagram of $\ung$. Let $\lambda\in P^+$,
$\bp\in\calp^{\lambda}$, and $\mu\in\lambda- Q_J^+$. Then, if $M$ is any
highest weight representation of $\uqgh$ with highest weight $\bp$ and highest
weight vector $m$, we have
$$m_{\mu}(M) =m_{\mu_J}(M_J),$$
where, $M_J=\uqgsh.m$.\endproclaim

\demo{Proof} If $V$ is any type 1 representation of $\uqgh$, and $\mu\in P$,
set
$$ V_{\mu}^+ =\{v\in V_{\mu}\mid x_{i,0}^+.v =0 \;{\text {for all}}\; i\in
I\}.$$
Similarly, if $W$ is any type 1 representation of $\uqgsh$, and $\nu\in P_J$,
define $W_{\nu}^+$ in the obvious way. It is clear that
$$m_{\mu}(M) = dim(M^+_{\mu}), \ \ \ \  m_{\mu_J}(M_J)=dim((M_J)^+_{\mu_J}).$$
Thus, it suffices to prove that
$$M_{\mu}^+ =(M_J)_{\mu_J}^+. \tag6$$
If $v\in M^+_{\mu}$, then, by 1.3(b), $v\in \hat{U}^-_{\lambda-\mu}.m$, where,
for any $\eta\in Q^+$,
$$\hat{U}^-_{\eta} =\{u\in \hat{U}^-\mid k_iuk_i^{-1} =q_i^{\eta(i)}u \;{\text{
for all}}\; i\in I\}.$$
Since $\lambda-\mu\in Q_J^+$, it follows that $v\in \hat{U}_J^-.m$, and hence
that $v\in(M_J)_{\mu_J}^+$. Conversely, since conjugation by $k_i $ clearly
preserves $(M_J)_{\mu_J}^+\subseteq M$ for all $i\in I$, it suffices to prove
that every $\uqg$--weight vector $v$ in $(M_J)_{\mu_J}^+$ belongs to
$M^+_{\mu}$.
If $v\in M_{\nu}$, then $\nu_J =\mu_J$, and $\nu\in\lambda-Q_J^+$ by 1.3(b).
This implies that $\nu =\mu$, since restriction to $J$ is injective on $Q_J^+$.
That $x_{i,0}^+.v =0$ for all $i\in I\backslash J$  is now clear, and the
converse is proved.\qed\enddemo

The assumption that $J$ is connected in 2.3 and 2.4 guaranteed that $\ung_J$
was simple, and hence standard results about $\uqg$ and $\uqgh$ could be
applied to $\uqgs$ and $\uqgsh$. The next two lemmas describe some consequences
of restricting to disconnected subdiagrams.
\proclaim{Lemma 2.5} Let $J_1,J_2\subseteq I$ be
non--empty subsets for which $a_{ij} =0$ if $i\in J_1$, $j\in J_2$ (in
particular, $J_1\cap J_2 =\emptyset$). Let $\lambda\in P^+$ and assume that
$\lambda_{J_2} =0$. If $\bp\in\calp^{\lambda}$ and $\mu$ is a weight of
$V(\bp)$ in $\lambda-Q_{J_1\cup J_2}^+$, then $\mu\in\lambda
-Q_{J_1}^+$.\endproclaim
\demo{Proof}  By 1.3, every vector in $V(\bp)_{\mu}$ is a linear combination of
vectors of the form
$$x_{i_1,r_1}^-x_{i_2,r_2}^-\ldots x_{i_k,r_k}^-.\vbp, \tag7$$
where $i_1,i_2,\ldots ,i_k\in J_1\cup J_2$, $r_1,r_2,\ldots, r_k\in\Bbb Z$,
$k\ge 1$. Since $a_{ij} =0$ if $i\in J_1$, $j\in J_2$, the relations in 1.2
tell us that
$$[x_{i,r}^-, x_{j,s}^-] =0 $$
if $i\in J_1$, $j\in J_2$, $r,s\in\Bbb Z$. Hence, we may assume that, in any
expression (7), all of the $x_{i,r}^-$'s with $i\in J_2$ occur to the right of
all $x_{i,r}^-$'s with $i\in J_1$. Since $\lambda_{J_2} =0$, it follows that
$x_{i,r}^-.\vbp =0$ if $i\in J_2$, $r\in\Bbb Z$, so an expression of type (7)
vanishes unless $i_1,\ldots,i_k$ all belong to $J_1$.\qed\enddemo

If $\emptyset\ne J\subseteq I$, $\lambda\in P$, let $\lambda^J\in P$ be defined
by
$$\lambda^J(i)=\cases
\lambda(i)& \text{if $i\in J$,}\\
0&\text{if $i\notin J$.}\endcases $$
Similarly, if $\bp= (P_i)_{i\in I}\in\calp$, let $\bp^J\in\calp$ have $i^{th}$
component equal to $P_i$ if $i\in J$, and equal to 1 otherwise.
\proclaim{Lemma 2.6} Let
$$I =J_1\amalg \{i_0\}\amalg J_2$$
(disjoint union), where $J_1$ and $J_2$ are such that  $a_{ij} =0$ if $i\in
J_1$, $j\in J_2$. Let $\lambda\in P^+$, $\bp\in\calp^{\lambda}$,  and let
$\mu\in P^+$ be of the form
$$\mu =\lambda -\sum_{j\in I, j\ne i_0}r_j\alpha_j, \ \ \ \ \  (r_j\in\Bbb
N).$$
Then, any $\uqg$--highest weight vector $v$ in $(V(\bp^{J_1\amalg\{i_0\}})\ot
V(\bp^{J_2}))_{\mu}$ (resp. in $(V(\bp^{J_1})\ot
V(\bp^{J_2\amalg\{i_0\}}))_{\mu}$) can be written
$$v = \sum_t w_t\ot w_t', \tag8$$
where $w_t\in V(\bp^{J_1\amalg\{i_0\}})$, $w_t'\in V(\bp^{J_2})$ (resp. $w_t\in
V(\bp^{J_1})$, $w_t'\in V(\bp^{J_2\amalg\{i_0\}})$), and $w_t$, $w_t'$ are
$\uqg$--highest weight vectors of weights $\lambda^{J_1\amalg\{i_0\}}
-\sum_{j\in J_1} r_j\alpha_j$ and $\lambda^{J_2} -\sum_{j\in J_2}r_j\alpha_j$
(resp. $\lambda^{J_1}-\sum_{j\in J_1}r_j\alpha_j$ and
$\lambda^{J_2\amalg\{i_0\}}-\sum_{j\in J_2}r_j\alpha_j$).
\endproclaim
\demo{Proof} We consider the tensor product $V(\bp^{J_1\amalg\{i_0\}})\ot
V(\bp^{J_2})$ (the proof in the other case is similar). We can obviously write
$v$ in the form (8) for some non--zero $\uqg$ weight vectors
$w_t$ and $w_t'$, of weights $\mu_t$ and $\mu_{t'}$, say. We may assume,
without loss of generality, that the $w_t'$ are linearly independent. Since
$\mu_t+\mu_t'=\mu$ for all $t$, it now follows from 2.5 that
$\mu_t\in\lambda^{J_1\amalg\{i_0\}}-Q_{J_1}^+$,
$\mu_t'\in\lambda^{J_2}-Q_{J_2}^+$. For weight reasons, it is clear that
$$x_j^+.w_t =0 \ \text{if $j\in J_2\amalg \{i_0\}$,}\ \ \ \ \ x_j^+.w_t' = 0 \
\text{if $j\in J_1\amalg \{i_0\}$.}$$
Hence, if $j\in J_1$, we have
$$ x_j^+.v =\sum_t(x_j^+.w_t\ot k_j.w_t' +w_t\ot x_j^+.w_t') = 0,$$
so
$$\sum_t q_j^{\mu_t'(j)} x_j^+.w_t\ot w_t' = 0.$$
Since the $w_t'$ are linearly independent, it follows that $x_j^+.w_t =0$ for
all $j\in J_1$. Hence, each $w_t$ is a $\uqg$--highest weight vector.
Interchanging the roles of $w_t$ and $w_t'$ one shows that the $w_t'$ are also
$\uqg$--highest weight vectors, thus proving the lemma.\qed
\enddemo

\vskip36pt\noindent{\bf 3 The $sl_{n+1}(\Bbb C)$ case}
\vskip12pt\noindent
If $\ung$ is of type $A_n$, we take $I =\{1,\ldots, n\}$, where $a_{ii} =2$,
$a_{ij} =-1$ if $|i-j|=1$, and $a_{ij}=0$ otherwise.
The following result describes the minimal affinizations of $\lambda$, for all
$\lambda\in P^+$, in this case.

By the $q$-segment of length $r\in\Bbb N$ and centre $a\in\Bbb C^\times$, we
mean the set of complex numbers $\{aq^{-r+1},aq^{-r+3},\ldots,aq^{r-1}\}$.
\proclaim{Theorem 3.1} Let $\ung=sl_{n+1}(\Bbb C)$, and let $\lambda\in P^+$.
Then, $\calq^{\lambda}$ has a unique minimal element. Moreover, this element is
represented by $V(\bp)$, for $\bp\in\calp^{\lambda}$, if and only if, for all
$i\in I$ such that $\lambda(i) >0$, the roots of $P_i$ form the $q$--segment
with centre $a_i$, for some $a_i\in\Bbb C^{\times}$, and length $\lambda(i)$,
where either

(a)  for all $i<j$, such that $\lambda(i) > 0$ and $\lambda(j) > 0$,
$$\frac{a_i}{a_j} = q^{ \lambda(i)+2(\lambda(i+1)+\cdots +\lambda(j-1))
+j-i},$$
or

(b)  for all $i<j$, such that $\lambda(i) > 0$ and $\lambda(j) > 0$,
$$\frac{a_j}{a_i} = q^{ \lambda(i)+2(\lambda(i+1)+\cdots +\lambda(j-1))
+j-i}.$$
In both cases, $V(\bp)\cong V(\lambda)$ as representations of
$\uqg$.\endproclaim
\demo{Proof} By Theorem 2.9 in [4], if $\bp\in\calp^{\lambda}$, then $V(\bp)$
is irreducible as a representation of $U_q(sl_{n+1})$ if and only if the
conditions in 3.1 hold. It is obvious that $[V(\bp)]$ is then the unique
minimal element of $\calq^{\lambda}$.\qed\enddemo
As an immediate consequence, we have
\proclaim{Corollary 3.2} Let $\ung =sl_{n+1}(\Bbb C)$, and let $\emptyset\ne
J\subseteq I$ define a connected subdiagram of the Dynkin diagram of $\ung$
(which is therefore of type $A_{|J|}$). Let   $\lambda\in P^+$ and
$\bp\in\calp^{\lambda}$ be such that $V(\bp)$ is a minimal affinization of
$\lambda$. Then:

(a) $V(\bp_J)$ is a minimal affinization of $\lambda_J$, and

(b) $V(\bp^J)$ is a minimal affinization of $\lambda^J$.\qed

\endproclaim
The following result is of crucial importance in the next section.
\proclaim{Proposition 3.3} Let $\ung =sl_{n+1}(\Bbb C)$, let $\lambda\in P^+$,
and let $\bp\in\calp^{\lambda}$ be such that

(a) $V(\bp)$ is not  a minimal affinization of $\lambda$, and

(b) $V(\bp_{I\backslash\{i\}})$ is a minimal affinization of
$\lambda_{I\backslash\{i\}}$, for $i=1,n$.

\noindent Then, $m_{\lambda-\theta}(V(\bp)) >0.$
\endproclaim
\demo{Proof} As a representation of $\uqg$, we have, by 1.5(a),
$$V(\bp)=V_0\op\bigoplus_t V_t, \tag9$$
where $V_0\cong V(\lambda)$, $V_t\cong V(\lambda-\eta_t)$, and $\eta_t\in Q^+$,
$\eta_t\ne 0$ (the $\eta_t$ are not necessarily distinct). Let $v_{\bp}^+$ be a
$\uqgh$--highest weight vector in $V(\bp)$, and $\vbp^-$ a $\uqgh$--lowest
weight vector. We claim that either $x_0^+.\vbp^+\notin V_0$ or
$x_0^-.\vbp^-\notin V_0$. Indeed, suppose the contrary and let $v\in V_0$.
Then,
$$v = x^-.\vbp^+ =x^+.\vbp^-,$$
for some $x^{{}\pm{}}\in U^{{}\pm{}}$. Since $[x_0^{{}\pm{}}, x^{{}\mp{}}] =0$
by the relations in 1.1, it follows that
$$x_0^{{}\pm{}}.v = x_0^{{}\pm{}}x^{{}\mp{}}.\vbp^{{}\pm{}} =
x^{{}\mp{}}x_0^{{}\pm{}}.\vbp^{{}\pm{}} \in x^{{}\mp{}}. V_0\subseteq V_0.$$
But, since $k_0$ acts on $V(\bp)$ as $(k_1k_2\ldots k_n)^{-1}$, the algebra of
operators on $V(\bp)$ defined by the action of $\uqgh$ is generated by the
action of $\uqg$ and $x_0^{{}\pm{}}$.  It follows that $V_0$ is a
$\uqgh$--subrepresentation of $V(\bp)$, and hence that $V(\bp) = V_0$,
contradicting 3.3(i).

Write $\vbp$ for $\vbp^+$ from now on, and assume, without loss of generality,
that $x_0^+.\vbp\notin V_0$. Then, $x_0^+.\vbp$ must have non--zero component,
with respect to the decomposition (9), in some $V_t$ with $\eta_t\ne 0$. Then,
$\eta_t\le\theta$, and it suffices to prove that $\eta_t =\theta$.

Suppose for a contradiction that $\eta_t<\theta$. Then,
$$\eta_t =\sum_{i=1}^nr_i\alpha_i,$$
where each $r_i =0$ or $1$, and at least one $r_i =0$. If $r_1 =0$ (resp. $r_n
=0$), applying 2.3 and 2.4 with $J =I\backslash\{1\}$ (resp. $J
=I\backslash\{n\}$) gives
$$m_{\lambda-\eta_t}(V(\bp)) = m_{(\lambda-\eta_t)_J} (V(\bp_J)),$$
which vanishes by 3.1 because $V(\bp_J)$ is a minimal affinization of
$\lambda_J$ by 3.2(b). But this is impossible, since
$m_{\lambda-\eta_t}(V(\bp)) > 0$.

Thus, $r_i =0$ for some $1 <i <n$. Let
$$J_1 =\{1,2,\ldots ,i-1\},\ \ \ \ \ \ J_2 =\{i+1,i+2,\ldots ,n\}.$$
By 2.6, any $\uqgh$--highest weight vector $v$ in $(V(\bp^{J_1\amalg\{i\}})\ot
V(\bp^{J_2}))_{\lambda-\eta_t}$ is of the form
$$v =\sum_rw_r\ot w_r',$$
where the $w_r$ and the $w_r'$ are $\uqg$--highest weight vectors of weights
$\lambda^{J_1\amalg\{i\}}-\sum_{j<i}r_j\alpha_j$ and $\lambda^{J_2}
-\sum_{j>i}r_j\alpha_j$, respectively. But, by 3.2(b) and 3.3(b), both
$V(\bp^{J_1\amalg\{i\}})$ and $V(\bp^{J_2})$ are minimal affinizations, so, by
3.1, we have $r_j =0$ for all $j<i$ and for all $j>i$. But then $\eta_t = 0$, a
contradiction.
\qed\enddemo
We  isolate the result in the $sl_2$ case; this was proved in [4].

\proclaim{Proposition 3.4} Let $\ung=sl_2(\Bbb C)$. For any $r\in\Bbb N$,
$\Cal{Q}^{r\lambda_1}$ has a unique minimal element. This element is
represented by $V(P)$, where $P$ is any polynomial of degree $r$ whose roots
form a $q$--segment. If $[W]\in\Cal{Q}^{r{\lambda_1}}$ is not minimal, then
$m_{(r-2)\lambda_1}\!(W) > 0$.\qed\endproclaim

\vskip36pt\noindent{\bf 4 The main reduction}
\vskip12pt\noindent
In this section, we continue to assume that $\ung$ is an arbitrary
finite-dimensional complex simple Lie algebra. We show (see Proposition 4.2)
that minimal affinizations remain minimal on restriction to certain \lq
admissible\rq\   subdiagrams  of the Dynkin diagram of $\ung$. To explain the
meaning of \lq admissible\rq, suppose temporarily that $\ung$ is of type $D$ or
$E$. Let $i_0\in I$ be the unique node of the Dynkin diagram of $\ung$ which is
linked to three  nodes other than itself. The set $I$ can then be written as a
disjoint union
$$I =I_1\amalg I_2\amalg I_3\amalg\{i_0\}$$
such that

(i) $I_r\cup\{i_0\}$ is of type $A$, for $r=1,2,3$,

(ii) for each $r=1,2,3$, there exists exactly one $i\in I_r$ such that
$a_{ii_0}\ne 0$, and

(iii) $a_{ij} =0$ if $i\in I_r$, $j\in I_s$, $r\ne s$.

\noindent Clearly, $I_1$, $I_2$, $I_3$ are uniquely determined, up to a
permutation.

\proclaim{Definition 4.1} Let $J$ be a non--empty subset of $I$. If $\ung$ is
not of type $D$ or $E$, $J$ is admissible iff $J$ is of type $A$. If $\ung$ is
of type $D$ or $E$, then $J$ is admissible iff
the following two conditions are satisfied:

(i) $J\subseteq I_r\cup\{i_0\}$ for some $r=1,2,3$, and

(ii) $J$ is connected (or, equivalently, $J$ is of type $A$).
\endproclaim
\proclaim{Proposition 4.2}  Let $J\subseteq I$ be admissible, let $\lambda\in
P^+$, and let $\bp =(P_i)_{i\in I}\in\calp^{\lambda}$.  If $V(\bp)$ is a
minimal affinization of $\lambda$, then $V(\bp_J)$ is a minimal affinization of
$\lambda_J$.
\endproclaim
\noindent{\it Remark.} This result is definitely false if $J$ is not
admissible, as will become clear in Theorem 6.1.
\demo{Proof of 4.2} The proof proceeds by induction on $|J|$. If $|J| =1$, we
must prove, in view of 3.4, that the roots of of each $P_i$ form a
$q_i$-segment.

Assume first that $i$ is linked to exactly one other node in $I$, and  suppose
for a contradiction that the roots of $P_i$ do not form a $q_i$--segement. Let
$Q_i$ be any polynomial with constant term 1 such that $deg(Q_i) =deg(P_i)$,
and whose roots do form a $q_i$-segment. Let $\bq$ be the $I$--tuple which is
equal to $\bp$ except in the $i^{th}$ place, where it equals $Q_i$. We prove
that $[V(\bq)]\prec [V(\bp)]$, giving the desired contradiction to the
minimality of $V(\bp)$.

Note that, by taking $\mu=\lambda-\alpha_i$, $J=\{i\}$ in 2.4, and using 2.3
and the second part of 3.4, it follows that
$$m_{\lambda-\alpha_i}(V(\bp)) > 0,\ \ \ \ \  m_{\lambda-\alpha_i}(V(\bq)) =
0.$$
Thus, $[V(\bp)]\ne [V(\bq)]$. To prove that $[V(\bq)]\prec [V(\bp)]$, we must
prove that, for all $\mu\in P$, either 1.8(i) or 1.8(ii) holds.
We may assume that $\mu=\lambda-\sum_{j\in I}s_j\alpha_j$, $s_j\ge 0$, since
otherwise $m_{\mu}(V(\bp)) = m_{\mu}(V(\bq)) =0$. Suppose first that $s_i >0$.
We have just shown that, if $\mu=\lambda-\alpha_i$, then 1.8(i) holds, while if
$\mu<\lambda-\alpha_i$, then 1.8(ii) holds with $\nu=\lambda-\alpha_i$. On the
other hand, if $s_i =0$, then applying 2.4 with $J =I\backslash\{i\}$, we have
$$m_{\mu}(V(\bp)) =m_{\mu_J}(V(\bp_J)) =m_{\mu_J}(V(\bq_J)) =m_{\mu}(V(\bq)),$$
and so 1.8(i) holds (note that $I\backslash\{i\}$ is connected because of our
assumption on $i$).

Suppose now that node $i$ is linked to two other nodes, and asssume for a
contradiction that the roots of $P_i$ do not form a $q_i$--segment. It is easy
to see that there exist subsets $J_1, J_2\subseteq I$ such that

(a) $I = J_1\amalg\{i\}\amalg J_2$ (disjoint union),

(b) $J_1\cup\{i\}$ defines a diagram of type $A$,

(c) $J_2$ is connected, and

(d) $a_{jk} =0$ if $j\in J_1$, $k\in J_2$.

\noindent Let $\bp'\in\calp^{\lambda_{J_1}\cup\{i\}}$ be such that $V(\bp')$ is
a minimal affinization of $\lambda_{J_1\cup\{i\}}$, and let $\bq=(Q_j)_{j\in
I}$ be defined by
$$Q_j= \cases
P_j & \text{if $j\in J_2$,}\\
P_j' & \text{if $j\in J_1\cup\{i\}$.}\endcases$$
We claim that $[V(\bq)]\prec [V(\bp)]$, giving a contradiction as before.

As in the first part of the proof, we see that $[V(\bq)]\ne [V(\bp)]$ and that,
in proving that $[V(\bq)]\preceq [V(\bp)]$, we need only consider weights
$\mu\in P$ of the form $\mu =\lambda-\sum_{j\in I}s_j\alpha_j$, where $s_j\ge
0$ for all $j\in I$, $s_i =0$, and $m_{\mu}(V(\bq)) > 0$. We show that, for
such $\mu$,
$$m_{\mu}(V(\bq)) =m_{\mu}(V(\bp)),$$
establishing 1.8(i) and proving our claim.

We make use of the following lemma, which will also be needed later.
\proclaim{Lemma 4.3} Let $i\in I$ be such that
$$I =J_1\amalg\{i\}\amalg J_2$$
(disjoint union), where $J_1$ is of type $A$, $J_2$ is connected, and $a_{jk}
=0$ if $j\in J_1$, $k\in J_2$. Let $\lambda\in P^+$, $\bq\in\calp^{\lambda}$,
and assume that $V(\bq^{J_1})$ is a minimal affinization of $\lambda^{J_1}$.
Let $\mu\in P$ be of the form $\mu =\lambda-\sum_{j\in I}s_j\alpha_j$, where
$s_j\ge 0$ for all $j$, and $s_{i} =0$. If $m_{\mu}(V(\bq)) > 0$, then $s_j =0$
for all $j\in J_1$ (and so $\mu\in\lambda - Q_{J_2}^+$).
\endproclaim

Assuming this lemma for the moment, we see that, if $m_{\mu}(V(\bq)) > 0$, then
$\mu\in\lambda-Q_{J_2}^+$. Since $\bp_{J_2} =\bq_{J_2}$, 2.4 implies, as
desired, that $m_{\mu}(V(\bp)) =m_{\mu}(V(\bq))$.

We have now proved 4.2 when $|J| =1$. For the inductive step, assume that $|J|
=r >1$ and suppose that the result is known when $|J| <r$. Proceeding by
contradiction, we suppose that $V(\bp_J)$ is a non--minimal affinization of
$\lambda_J$. Define a subset $J'\subseteq I$ and a node $j_0\in J$ as follows:
\vskip6pt
(i) if $J$ contains an element $j$ that is linked to exactly one other element
in $I$, choose $j_0=j$ and $J'=\emptyset$;

(ii) otherwise, choose $J'$ to be disjoint from $J$ such that $J\cup J'$ is
admissible and $I\backslash (J\cup J')$ is connected, and let $j_0$ be the
unique element of $J$ that is connected to an element of $J'$.
\vskip6pt
\noindent See the diagrams on the next page.

By the induction hypothesis, $V(\bp_{J\backslash\{j_0\}})$ is a minimal
affinization. Hence, by 3.1, we may choose $P_j'$, for $j\in J'\cup\{j_0\}$,
such that $deg(P_j) =deg(P_j')$, and such that, if we define the  $(J\cup
J')$--tuple $\bold R=(R_j)_{j\in J\cup J'}$ by
$$R_j=\cases P_j & \text{if $j\in J\backslash\{j_0\}$,}\\
P_j' & \text{if $j\in J'\cup\{j_0\}$,}\endcases$$
then $V(\bold R)$ is a minimal affinization of $\lambda_{J\cup J'}$. Now define
$\bq= (Q_j)_{j\in I}\in\calp^{\lambda}$ by
$$Q_j=\cases
 P_j'& \text{if $j\in J'\cup\{j_0\}$,}\\
P_j &\text{otherwise}.\endcases$$
We prove that $[V(\bq)] \prec [V(\bp)]$, giving the usual contradiction.

Note first that, by 3.2, $V(\bq_J)$ is a minimal affinization of $\lambda_J$,
but by assumption, $V(\bp_J)$ is not minimal. By 3.3,
$$m_{\lambda_J -\sum_{i\in J}(\alpha_i)_J}(V(\bp_J)) > 0, \ \ \ \ \
m_{\lambda_J-\sum_{i\in J}(\alpha_i)_J}(V(\bq_J)) = 0.$$
By 2.3 and 2.4,
$$m_{\lambda-\sum_{i\in J}\alpha_i}(V(\bp)) > 0,\ \ \ \ \
m_{\lambda-\sum_{i\in J}\alpha_i}(V(\bq)) = 0.\tag10$$
Hence, $[V(\bp)] \ne [V(\bq)]$.

\topspace{200pt}

To prove that $[V(\bq)]\prec [V(\bp)],$ we need only consider, as usual,
weights $\mu$ such that $m_{\mu}(V(\bq)) > 0$ and $\mu=\lambda-\eta$, where
$\eta =\sum_{j\in I}s_j\alpha_j$ and each $s_j\ge 0$. By the second equation in
(10), $\eta\ne\sum_{j\in J}\alpha_j$. If $\eta >\sum_{j\in J}\alpha_j$, then
1.8(ii) holds with $\nu=\lambda-\sum_{j\in J}\alpha_j$. Hence, we may assume
that $s_{j_1} =0$ for some $j_1\in J$. Define a subset  $J''$ of $J$ as
follows:
\vskip6pt
(i) $J'' =J$ if $j_0 =j_1$,

(ii) if $j_0\ne j_1$, then $J''$ is the maximal connected subset of $J$
containing $j_0$ but not $j_1$.
\vskip6pt
\noindent See the diagrams on the next page.

Set $J_1 =J'\cup J''$, $J_2 =I\backslash (J_1\cup\{j_1\})$. Note that $J_1$ is
of type $A$ and $J_2$ is connected. Applying 4.3, we see that $\mu\in\lambda
-Q_{J_2}^+$. Since $\bp_{J_2} = \bq_{J_2}$ it follows as usual from 2.3 and 2.4
that
$$m_{\mu}(V(\bp)) = m_{\mu}(V(\bq)),$$
 thus completing the proof of the inductive step.\qed\enddemo

All that remains is to give the
\demo{Proof of 4.3} By 1.6(d), $V(\bq)$ is isomorphic to a subquotient of the
tensor product $V(\bq^{J_1})\ot V(\bq^{J_2\cup\{i_0\}})$; a fortiori,
$m_{\mu}(V(\bq^{J_1})\ot V(\bq^{J_2\cup\{i_0\}}) )> 0$. By 2.6, if $v\in
(V(\bq^{J_1})\ot V(\bq^{J_2\cup\{i_0\}}))_{\mu}$ is any $\uqg$--highest weight
vector, then
$$v =\sum_t w_t\ot w_t',$$
where $w_t\in V(\bq^{J_1})$ is a $\uqg$--highest weight vector of weight
$\lambda^{J_1} -\sum_{j\in J_1}s_j\alpha_j$, and $w_t'\in
V(\bq^{J_2\cup\{j_0\}})$ is a $\uqg$--highest weight vector of weight
$\lambda^{J_2\cup\{j_0\}} -\sum_{j\in J_2} s_j\alpha_j$. Since $V(\bq^{J_1})$
is a minimal affinization of $\lambda^{J_1}$, 3.1 implies that $s_j =0$ for all
$j\in J_1$ and hence $\mu=\lambda- Q_{J_2}^+$.
\qed\enddemo

\topspace{200pt}

\vskip36pt\noindent
{\bf 5 Twisting with the Cartan involution}
\vskip12pt\noindent
In this section, $\ung$ is an arbitrary finite-dimensional complex simple Lie
algebra. If $V$ is any representation of $\uqgh$, given by a homomorphism
$\rho:\uqgh\to End(V)$, say, we denote by $\hat{\omega}^*(V)$ the
representation $\rho\circ\hat{\omega}$, where $\hat{\omega}$ is the involution
of $\uqgh$ defined in 1.4(b).  Let $V^*$ be the $\uqgh$--representation dual to
$V$: recall that the action of $\uqgh$ on $V^*$ is defined by

$$(x.f)(v) = f(S(x). v),$$
where $f\in V^*$, $x\in\uqgh$, and $S: \uqgh\to\uqgh$ is the antipode.  It is
clear that, if $V$ is an irreducible representation of $\uqgh$, then $V^*$ and
$\hat{\omega}^*(V)$ are both irreducible representations as well. The purpose
of this section is to give the defining polynomials of $\hat{\omega}^*(V)$  and
$V^*$ in terms of the defining polynomials of $V$. We need this result in the
next section to prove the uniqueness of certain minimal affinizations.

Let $w_0$ be the longest element of the Weyl group of $\ung$, and let
$i\to\overline{i}$ be the bijection $I\to I$ such that $w_0(\alpha_i)
=-\alpha_{\overline{i}}$. It is well known that
$${\omega}^*(V(\lambda))\cong V(-w_0(\lambda)),\ \ \ \ \ \ V(\lambda)^* \cong
V(-w_0(\lambda)),$$
for all $\lambda\in P^+$, where $\omega$ is the Cartan involution of $\uqg$.
\proclaim{Proposition 5.1} Let $\lambda\in P^+$, $\bp =(P_i)_{i\in
I}\in\calp^{\lambda}$, and let
$$P_i(u) =\prod_{r=1}^{\lambda(i)}(1-a_{r,i}^{-1}u), \ \ \ \ \ \ \
(a_{r,i}\in\Bbb C^{\times}).$$

(a) Define ${\bp}^{\hat{\omega}} =(P^{\hat{\omega}}_i)_{i\in
I}\in\calp^{-w_0(\lambda)}$ by
$$P^{\hat{\omega}}_{\overline{i}}(u)
=\prod_{r=1}^{\lambda(i)}(1-q_i^2a_{r,i}u).$$
Then, there exists $t\in\Bbb C^{\times}$, independent of $i\in I$, such that
$$\hat{\omega}^*(V(\bp))\cong \tau_{t}^*(V({\bp}^{\hat{\omega}}))$$
as representations of $\uqgh$.

(b) Define $\bp^* = (P_i^*)_{i\in I}\in\calp^{-w_0(\lambda)}$ by
$${P}^*_{\overline{i}}(u) =\prod_{r=1}^{\lambda(i)}(1-a_{r,i}^{-1}u).$$
Then, there exists $t^*\in\Bbb C^{\times}$ such that,
as representations of $\uqgh$,
$$V(\bp)^*\cong \tau_{t^*}^*(V(\bp^*)).$$

\endproclaim

\demo{Proof} We first prove that it suffices to establish the proposition in
the case when $\lambda$ is fundamental. We do this for part (b); the proof for
part (a) is similar (see also [2], where the corresponding result was proved
for rank two algebras).. By 1.6(d), we see that $V(\bp)$ is the unique
irreducible subquotient of
$$\bigotimes_{i\in I}\bigotimes_{r=1}^{\lambda(i)}V(\lambda_i,a_{i,r})$$
which contains the tensor product of the highest weight vectors (the tensor
product of the representations can be taken in any order). It is not hard to
see that $$V(\lambda_i, a_{i,r})^*\cong  V(\lambda_{\overline{i}},
a_{i,r}^*),$$
for some $a_{i,r}^*\in \Bbb C^{\times}$ (this follows from Proposition 3.3 in
[2]). Hence, $V(\bp)^*$ is the unique irreducible subquotient of
$$\bigotimes_{i\in I}\bigotimes_{r=1}^{\lambda(i)}
V(\lambda_{\overline{i}},a_{i, r}^*)$$
containing the tensor product of the highest weight vectors. Thus, by 1.6(d),
it suffices to calculate the $a_{i,r}^*$.

The proof of 5.1 in the  fundamental case is a consequence of the following
lemma.

\proclaim{Lemma 5.2} Suppose that $a_{ij}\ne 0$, $i\ne j$, and that $a_i,
a_j\in\Bbb C^{\times}$. Then,

(a) $m_{\lambda_i+\lambda_j-\alpha_i-\alpha_j}(V(\lambda_i,a_i)\ot
V(\lambda_j,a_j)) =1$;

(b) if $v_i\in V(\lambda_i,a_i)$, $v_j\in V(\lambda_j,a_j)$ are
$\uqgh$--highest weight vectors, and $M =\uqgh.(v_i\ot v_j)\subset
V(\lambda_i,a_i)\ot V(\lambda_j,a_j)$, then
$m_{\lambda_i+\lambda_j-\alpha_i-\alpha_j}(M) =0$ iff $$\frac{a_i}{a_j}
=q^{-(3d_i+d_j-1)};$$

(c) Let $v\in (V(\lambda_i,a_i)\ot V(\lambda_j ,a_j))
_{\lambda_i+\lambda_j-\alpha_i-\alpha_j}$ be a $\uqg$--highest weight vector.
Then, $v$ is also $\uqgh$--highest weight iff $$\frac{a_i}{a_j}
=q^{3d_j+d_i-1}.$$\endproclaim

Assuming this lemma, 5.1(a) is proved as follows. Using the notation introduced
in 5.2, we have
$$\hat{\omega}^*(M)\subseteq \hat{\omega}^*(V(\lambda_j,a_j))\ot
\hat{\omega}^*(V(\lambda_i,a_i)).$$
As in the proof of Proposition 5.5 in [2],
$$\hat{\omega}^*(V(\lambda_i,a_i))\cong V(\lambda_{\overline i},
\overline{a}_i)$$
for some $\overline{a}_i\in\Bbb C^{\times}$. Identifying the two
representations above, we thus have
$$\hat{\omega}^*(M)\subseteq V(\lambda_{\overline{j}},\overline{a}_j)\ot
V(\lambda_{\overline{i}},\overline{a}_i).$$
Now, since $m_{\lambda_i+\lambda_j}(M) =1$, we have
$m_{\lambda_{\overline{i}}+\lambda_{\overline{j}}}(\hat{\omega}^*(M)) =1$
by the discussion preceding 5.1. Hence, $\hat{\omega}^*(M)$ contains
$\uqgh.(v_{\overline{j}}\ot v_{\overline{i}})\subseteq
V(\lambda_{\overline{j}}, \overline{a}_j)\ot V(\lambda_{\overline{i}},
\overline{a}_i)$.
Assume now that ${a_i}/{a_j} =q^{-(3d_i+d_j-1)}$. Then, by 5.2(b),
$m_{\lambda_i+\lambda_j-\alpha_i-\alpha_j}(M) =0$, hence $m_{\lambda_{\overline
i}+\lambda_{\overline j}-\alpha_{\overline i}-\alpha_{\overline
j}}(\hat{\omega}^*(M)) = 0$. A fortiori,
 $m_{\lambda_{\overline i}+\lambda_{\overline j}-\alpha_{\overline
i}-\alpha_{\overline j}}(\uqgh.(v_{\overline j}\ot v_{\overline i})) =0$. By
5.2(b) again, ${\overline a_j}/{\overline a_i} = q^{-(3d_{\overline
j}+d_{\overline i} -1)}$. Since $d_{\overline i} = d_i$ for all $i\in I$, we
get
$$q_j^2\overline{a}_ja_j=q_i^2\overline{a}_ia_i,$$
from which 5.1(a) follows for fundamental representations.

We now prove 5.1(b). We continue to assume that
$$\frac{a_i}{a_j} =q^{-(3d_i+d_j-1)}.$$
Let $a_i^*\in\Bbb C^\times$ be such that
$$V(\lambda_i, a_i)^* = V(\lambda_{\overline{i}}, a_i^*).$$
By standard properties of duals, $M^*$ is a quotient of
$V(\lambda_{\overline{j}},a_j^*)\ot V(\lambda_{\overline i},a_i^*)$. Since
$m_{\lambda_i+\lambda_j-\alpha_i-\alpha_j}(M) =0$, we have
$m_{\lambda_{\overline{i}}+\lambda_{\overline{j}}-\alpha_{\overline{i}}-
\alpha_{\overline{j}}}(M^*) =0$. Applying 5.2(c), we see that
$$\frac{a_j^*}{a_i^*}
=q^{3d_{\overline{i}}+d_{\overline{j}}-1}=q^{3d_i+d_j-1}.$$
This gives
$$\frac{a_i}{a_i^*} =\frac{a_j}{a_j^*},$$
from which 5.1(b) follows.
\qed\enddemo

\demo{Proof of 5.2(a)} It suffices to prove that, if $a_{ij}\ne 0$, $i\ne j$,
and $a_i\in\Bbb C^{\times}$, then
$$m_{\lambda_i-\alpha_i}(V(\lambda_i,a_i))
=m_{\lambda_i-\alpha_i-\alpha_j}(V(\lambda_i,a_i)) = 0.\tag11$$
For, this result clearly implies that
$$m_{\lambda_i+\lambda_j-\alpha_i-\alpha_j}(V(\lambda_i,a_i)\ot
V(\lambda_j,a_j))
=m_{\lambda_i+\lambda_j-\alpha_i-\alpha_j}(V(\lambda_i)\ot V(\lambda_j)),$$
and it easy to see that the last multiplicity is one.

It suffices to prove (11) when $\ung$ is of rank 2. For, if $J=\{i,j\}\subseteq
I$, then, by the rank 2 case,
$m_{(\lambda_i-\alpha_i-\alpha_j)_J}(V((\lambda_i)_J,a_i)) =0$, so, by 2.4,
$m_{\lambda_i-\alpha_i-\alpha_j}(V(\lambda_i,a_i)) =0$.

If $\ung$ is of type $A_2$, (11) is obvious, since, by 3.1, $V(\lambda_i,a_i)$
is an irreducible representation of $\uqg$.

If $\ung$ is of type $C_2$ or $G_2$, this was proved in [2], Proposition
5.4(i).
\qed\enddemo

\demo{Proof of 5.2(b), (c)} Taking $J =\{i,j\}$ we see that, by Proposition
2.2,  it suffices to prove this result in the rank two case. If $\ung_J$ is of
type $A_2$, both parts (b) and (c) are established in the proof of Lemma 4.1 in
[4].

If $\ung$ is of type $C_2$ or $G_2$, then $i=\overline{i}$ for $i=1,2$. Part
(b) was established in Proposition 5.4(c) in [2]. To prove (c), notice that, by
(a),  $v$ is a $\uqgh$--highest weight vector in $V(\lambda_i,a_i)\ot
V(\lambda_j,a_j)$ iff $m_{\lambda_i+\lambda_j-\alpha_i-\alpha_j}(M^*) =0$,
where $M^* =\uqgh(v_j\ot v_i) \subseteq V(\lambda_j,a_j)^*\ot
V(\lambda_i,a_i)^*$.
Writing $V(\lambda_i,a_i)^*\cong V(\lambda_i,a_i^*)$, we see from part (b) that
$$\frac{a_j^*}{a_i^*} =q^{-(3d_j+d_i-1)}.\tag12$$
A direct calculation in the rank two case now gives that $$a_r^* =ta_r,$$ and
combining with (12) gives the desired result.

\enddemo

\vskip36pt\noindent
{\bf 6 The simply--laced case}
\vskip12pt\noindent
In this section, we assume that $\ung$ is of type $D$ or $E$. Let
$I_1,I_2,I_3\subset I$, and $i_0\in I$, be as defined at the beginning of
Section 4. If $\lambda\in P$, define subsets $I_r(\lambda)\subseteq I_r$,
$r=1,2,3$, by the following conditions:

(i) $\lambda_{I_r(\lambda)} =0$,

(ii) $I_r(\lambda)$ is connected,

(iii) $I_r(\lambda)\cup\{i_0\}$ is of type $A$, and

(iv) $I_r(\lambda)$ is maximal with respect to properties (i)--(iii).

\noindent Note that $I_r(\lambda)$ may be empty.

The following theorem is the main result of this paper.

\proclaim{Theorem 6.1} Let $\ung$ be of type $D$ or $E$. Let $\lambda\in P^+$
and assume that $\lambda(i_0) \ne 0$.

(a) If $I_r(\lambda) =I_r$ for some $r\in\{1,2,3\}$, then $\calq^{\lambda}$ has
a unique minimal element. This element is represented by $V(\bp)$, where
$\bp\in\calp^{\lambda}$, if and only if $V(\bp_{I\backslash I_r})$  is a
minimal affinization of $\lambda_{I\backslash I_r}$.

(b) Suppose that, for all $r\in\{1,2,3\}$, $I_r(\lambda)\ne I_r$. Then,
$\calq^{\lambda}$ has exactly three minimal elements. In fact, if
$\bp\in\calp^{\lambda}$, then $[V(\bp)]$ is minimal if and only if there exists
$r,s\in\{1,2,3\}$, $r\ne s$, such that $V(\bp_{I\backslash I_r})$ and
$V(\bp_{I\backslash I_s})$ are minimal affinizations of $\lambda_{I\backslash
I_r}$ and $\lambda_{I\backslash I_s}$, respectively.
\endproclaim
\vskip6pt\noindent
{\it Remarks.} 1. Note that, for any $r\in\{1,2,3\}$, $I\backslash I_r$ is of
type $A$, so we know from the results of Section 3 precisely when
$V(\bp_{I\backslash I_r})$ is minimal.

2. It might be helpful to illustrate this theorem diagrammatically. First, if
$\ung$ is of type $A$, $\lambda\in P^+$,
$\bp=(P_i)_{i\in I}\in{\Cal P}^\lambda$, and if the roots of $P_i$ form a
$q$-segment with centre $a_i$ for all $i\in I$, then we draw an arrow above the
Dynkin diagram of $\ung$
\vskip 1cm\centerline{or}\vskip1cm\noindent
according as the $a_i$ satisfy condition (a) or condition (b) in 3.1,
respectively. If $\ung$ is of type $D$ or $E$, the theorem says that, under the
hypotheses of 6.1(a), the minimal element of ${\Cal Q}^\lambda$ is given by the
diagram
\vskip4cm\noindent
and under the hypotheses of 6.1(b), the three minimal elements of
${\Cal Q}^\lambda$ are given by the diagrams
\vskip4cm

\demo{Proof of 6.1}  Suppose first that $I_r(\lambda)=I_r$ for all $r$. Then,
if $V(\bp)$ is minimal, by 4.2 the roots of $P_{i_0}$ form a
$q_{i_0}$--segment, and obviously $P_i =1$ if $i\ne i_0$. By 1.7, $V(\bp)$ is
unique up to twisting with an automorphism $\tau_t$, for some $t\in\Bbb
C^{\times}$. In particular, the element $[V(\bp)]\in\calq^{\lambda}$ is unique
and part (a) is proved in this case.

Suppose next that $I_r(\lambda) = I_r$ for exactly two values of $r$, say
$r=1,2$, without loss of generality. If $V(\bp)$ is a minimal affinization of
$\lambda$, then, by 4.2, $V(\bp_{I_3\cup\{i_0\}}\!\!)$ is a minimal
affinization of $\lambda_{I_3\cup\{i_0\}}$. By 3.1, for all $i\in
I_3\cup\{i_0\}$ such that $\lambda(i) > 0$, the roots of $P_i$ form a
$q_i$--segment with centre $a_i$, say, where $a_i/a_{i_0}$ satisfies either
condition (a)  or condition (b) in 3.1. By 5.1, $V(\bp)$ satisfies condition
(a) iff $(\hat{\omega}^*(V(\bp)))^*$ satisfies condition (b).  Since $[V(\bp)]
=[\hat{\omega}^*(V(\bp))^*]$ it follows that the equivalence class of $V(\bp)$
is uniquely determined.

For the remainder of the proof of 6.1(a), and also for the proof of 6.1(b), we
introduce the following notation. If $r\in\{1,2,3\}$, let $i_r\in I_r$ be the
unique index such that

(i) $\lambda(i_r)\ne 0$, and

(ii) $\{i_r\}\cup\{i_0\}\cup I_r(\lambda)$ is of type $A$.

\noindent Note that, if $I_r(\lambda)\ne I_r$, then $i_r$ and $i_0$ are the
nodes of $\{i_r\}\cup\{i_0\}\cup I_r(\lambda)$  which are connected to only one
other node (and $ i_r =i_0$ if $I_r(\lambda) =I_r$).

\vskip6cm\noindent

Define $\theta_r(\lambda) =\sum_{i\in I_r(\lambda)}\alpha_i\in Q^+$.

\proclaim {Proposition 6.2}
Let $\lambda\in P^+$ satisfy $\lambda(i_0) >0$, and let
$\bp\in\calp^{\lambda}$. Assume
 that $V(\bp_{I_r\cup\{i_0\}})$ is minimal for $r=1,2,3$.

\vskip 6pt\noindent
(i)  Let $\{r,s,t\} =\{1,2,3\}$. The following statements are equivalent:

(a) $V(\bp_{I\backslash I_r}\!\!\!)$ is a minimal affinization of
$\lambda_{I\backslash I_r}$;

(b) $V(\bp_{I_{s}(\lambda)\cup \{i_0, i_s, i_t\}\cup I_{t}(\lambda)})$ is a
minimal affinization of $\lambda_{I_{s}(\lambda)\cup \{i_0, i_s, i_t\}\cup
I_{t}(\lambda)}$;

(c) $m_{\lambda-\alpha_{i_0}-\alpha_{i_s}-\alpha_{i_t}-
\theta_s(\lambda)-\theta_t(\lambda)}(V(\bp)) = 0$.

\vskip6pt\noindent
(ii) Let $0\ne\eta=\sum_js_j\alpha_j\in Q^+$ be such that
$m_{\lambda-\eta}(V(\bp)) >0 $.  Then,

(a) $s_{i_0}\ne 0$;

(b) if $j\in I_r$ is such that $s_j >0$, and if $J\subseteq I_r\cup\{i_0\}$ is
the connected subset of type $A$ which has $j$ and $i_0$ as its \lq end\rq\
nodes, then $s_i >0$ for all $i\in J$;

(c) if $I_r\ne I_r(\lambda)$ then either
$s_j >0$ for all $j\in I_r(\lambda)$ or $s_j =0$ for all $j\in I_r\backslash
I_r(\lambda)$.

\endproclaim

\demo{Proof of 6.2} (i) The equivalence (a) $\Leftrightarrow$ (b)  is obvious
from 3.1. The equivalence (b) $\Leftrightarrow$ (c) follows from 2.4 and 3.3.

(ii) Suppose that $m_{\mu}(V(\bp)) > 0$. Write $\mu =\lambda-\eta$, where
$\eta=\sum_js_j\alpha_j$. Suppose that $s_{i_0} =0$. Let $\{r,s,t\}
=\{1,2,3\}$. Since $V(P_{I_r\cup\{i_0\}})$ is minimal of type A, it follows
from 2.4 and 3.1 that $m_{\nu}(V(\bp_{I_r\cup\{i_0\}}) = 0$ where
$\nu=\lambda_{I_r\cup\{i_0\}}-\eta'$,  and $\eta'\in Q^+_{I_r\cup\{i_0\}}$.
Applying 2.6 to the decomposition $I =I_r\cup\{i_0\}\cup(I_s\cup I_t)$ now
shows that $s_i =0$ for all $i\in I_r$, $r=1,2,3$. Hence, $\eta =0$,
contradicting our assumption. This proves  (a).

Let $j\in I_r$ be such that $s_{j} > 0$ and let $J\subseteq I_r\cup\{i_0\}$ be
the type $A$ subset which has $j$ and $i_0$ as its \lq end\rq\  nodes.
Suppose that  $s_i =0$ for some $i\in J$, say $i=j'$. We have a unique
decomposition
$$I = J'\amalg\{j'\}\amalg J''$$
(disjoint union), where $j\in J'\subset I_r$, $i_0\in J''\cup\{j'\}$,
$J'$ is of type $A$ and $a_{rs} =0$ if $r\in J'$, $s\in J''$.
Applying 2.6, 2.4 and 3.1  again gives that $s_i =0$ for all $i\in J'$,
contradicting
$s_j\ne 0$. This proves  (b).

Part (c) now follows by considering separately the cases $s_{i_r} >0$ and
$s_{i_r} =0$. $\qed$

\enddemo

We now return to the proof of 6.1(a) in the case  $I_1(\lambda) = I_1$,
$I_r(\lambda)\ne I_r$, $r=2,3$. Suppose for a contradiction that
$V(\bp_{I\backslash I_1})$ is not minimal. By 6.2(i) this means that
$$m_{\lambda-\theta_2(\lambda)-\theta_3(\lambda)-\alpha_{i_2}-\alpha_{i_3}-\alpha_{i_0}}(V(\bp)) >0.\tag13$$
By 3.1, there exists a unique $\bq =(Q_i)_{i\in I}\in\calp^{\lambda}$ such that

(i) $Q_i =1$ if $i\in I_1$;

(ii) $Q_i =P_i$ if $i\in I_2\cup\{i_0\}$;

(iii)$ V(\bq_{I\backslash I_1})$ is a minimal affinization of
$\lambda_{I\backslash I_1}$.

\noindent We prove that $[V(\bq)]\prec [V(\bp)]$, contradicting the minimality
of $[V(\bp)]$.

Clearly, $[V(\bq)]\ne [V(\bp)]$, since, by 6.2(i),
$$m_{\lambda-\theta_2(\lambda)-\theta_3(\lambda)-\alpha_{i_2}-\alpha_{i_3}-
\alpha_{i_0}}(V(\bq)) =0.\tag14$$

Suppose that $\mu\in P^+$ is such that $m_{\mu}(V(\bq)) >0$, and let
$\mu=\lambda-\eta$, $\eta\in Q^+$.
Write $\eta =\sum_js_j\alpha_j$. If $s_{i_2} >0$ and $s_{i_3} >0$, it follows
from 6.2(ii)(a) that $\eta > \theta_2(\lambda)
+\theta_3(\lambda)+\alpha_{i_0}+\alpha_{i_2}+\alpha_{i_3}$. Equations (13) and
(14) now show that
 condition 1.8(ii) is satisfied with
$\nu=\lambda-\theta_2(\lambda)-\theta_3(\lambda)-\alpha_{i_2}-\alpha_{i_3}-
\alpha_{i_0}$.

If $s_{i_2} \ge 0$ and $s_{i_3} =0$,
 let $J=I_1\cup I_2\cup I_3(\lambda)\cup\{i_0\}$. By 2.4 and the fact that
$\bp_J =\bq_J$, we get
$$m_{\mu}(V(\bq)) = m_{\mu_J}(V(\bq_J)) =m_{\mu_J}(V(\bp_J)) =
m_{\mu}(V(\bp)),$$
so 1.8(i) is satisfied. If $s_{i_2} =0$ and $s_{i_3}>0$, let $J' =I_1\cup
I_2(\lambda)\cup I_3\cup\{i_0\}$. By 2.4, it suffices to show that
$m_{\mu_{J'}}(V(\bp_{J'})) =m_{\mu_{J'}}(V(\bq_{J'}))$. Note that
$P_i =Q_i =1$ if $i\in J'\backslash (I_3\cup\{i_0\})$, and that, if $i\in
I_3\cup\{i_0\}$, then, by 4.2 and  3.1, there exists $a_i$, $\gamma\in \Bbb
C^{\times}$ such that the roots of $P_i$ (resp. $Q_i$) form a $q_i$--segment
with centre $a_i$ (resp. $\gamma a_i^{-1}$). It follows from 5.1 that
$$(\hat{\omega}^*(V(\bp_{J'})))^*\cong \tau_t^*(V(\bq_{J'}))$$
for some $t\in\Bbb C^{\times}$ (here, $\hat{\omega} $ and $\tau_t$ are the
appropriate automorphisms of $U_q(\hat{\ung}_{J'})$). This proves our
assertion.

We have now shown that, if $V(\bp)$ is a minimal affinization of $\lambda$,
then $V(\bp_{I\backslash I_1})$ is a minimal affinization of
$\lambda_{I\backslash I_1}$. Conversely, suppose that $V(\bp_{I\backslash
I_1})$ is minimal. Choose $\bq\in\calp^{\lambda}$ such that $V(\bq)$ is minimal
and $[V(\bq)]\preceq [V(\bp)]$. By the first part of the proof,
$V(\bq_{I\backslash I_1})$ is minimal. By 3.1, there exists $\gamma\in\Bbb
C^{\times}$ such that either
\vskip6pt
(i) for all $i\in I\backslash I_1$, there exists $a_i\in\Bbb C^{\times}$ such
that the roots of $P_i$ (resp. $Q_i$) form a $q_i$--segment with centre $a_i$
(resp. $\gamma a_i$),

\noindent or

(ii) for all $i\in I\backslash I_1$, there exists $a_i\in\Bbb C^{\times}$ such
that the roots of $P_i$ (resp. $Q_i$) form a $q_i$--segment with centre $a_i$
(resp. $\gamma a_i^{-1}$).
\vskip6pt
\noindent Since $P_i =Q_i =1$ for $i\in I_1$, it follows from 1.7 and 5.1 that
either
\vskip6pt
(i) $V(\bp) \cong\tau_t^*(V(\bq))$,

\noindent or

(ii) $V(\bp)\cong (\hat{\omega}^*\tau_t^*(V(\bq)))^*$,
\vskip6pt
\noindent for some $t\in\Bbb C^{\times}$. But, in both cases, $[V(\bp)]
=[V(\bq)]$, so $[V(\bp)]$ is minimal.

This completes the proof of 6.1(a).

Suppose now, for 6.1(b), that $I_r(\lambda)\ne I_r$ for all $r$, that $V(\bp)$
is a minimal affinization of $\lambda$, but that neither
$V(\bold{P}_{I\backslash I_2})$ nor $V(\bp_{I\backslash I_3})$ is minimal. By
3.1 and 4.2, it follows that $V(\bold{P}_{I\backslash I_1})$ is not minimal
either (this is clear from the diagrams in the second remark following the
statement of 6.1). By 6.2(i),
$$m_{\lambda-\alpha_{i_0}-\alpha_{i_r}-\alpha_{i_s}-\theta_r(\lambda)-\theta_s(\lambda)}(V(\bp )) > 0\tag15$$
for all $r\ne s$ in $\{1,2,3\}$. By 3.1 again, there exists
$\bq\in\calp^{\lambda}$ such that $V(\bq_{I\backslash I_1})$ is minimal and
$Q_i =P_i$ if $i\in I_1\cup I_2$. Notice that then $V(\bq_{I\backslash I_2})$
is also a minimal affinization of $\lambda_{I\backslash I_2(\lambda)}$ and
hence by 6.2(i)

$$m_{\lambda-\alpha_{i_0}-\alpha_{i_r}-\alpha_{i_3}-\theta_r(\lambda)
-\theta_3(\lambda)}(V(\bq)) = 0, \ \ \ r=1,2\tag16$$
By (15), $[V(\bq)]\ne [V(\bp)]$
and we prove next that $[V(\bp)]\prec [V(\bq)]$.

Suppose that $\mu=\lambda-\eta$, where $\eta\in Q^+$ is such that
$m_{\mu}(V(\bq)) > 0$.

If either $s_{i_1}, s_{i_3}> 0$ or $s_{i_2}, s_{i_3} >0$, then 6.2(ii)(a),
together with equations (15) and (16), shows  that
1.8(ii) holds with
$$\nu=\lambda-\alpha_{i_0}-\alpha_{i_r}-\alpha_{i_3}-\theta_r(\lambda)-\theta_3(\lambda)$$
for $r=1$ or $2$.

If $s_{i_3} =0$, set $J =I_1\cup I_2\cup\{i_0\}\cup I_3(\lambda)$. By
6.2(ii)(b), $\eta\in Q_J^+$, and noting that $\bp_J =\bq_J$, we get from 2.4
that
$$m_{\mu}(V(\bp)) =m_{\mu}(V(\bp_J)) =m_{\mu}(V(\bq_J))=m_{\mu}(V(\bq)).$$

Finally if $s_{i_3} >0$ and $s_{i_1}=s_{i_2}=0$, take $J= I_1(\lambda)\cup
I_2(\lambda)\cup\{i_0\}\cup I_3$. By 6.2(ii)(b), $\eta\in Q_J^+$.
The same argument used in this case in 6.1(a) shows that, for some
$\gamma\in\Bbb C^{\times}$,
$$V(\bp_J)\cong \tau^*_{\gamma}(V(\bq_J)) \ \ \ \  \text{or}\ \ \ \
V(\bp_J')\cong (\hat{\omega}^*\tau^*_{\gamma}(V(\bq_J)))^*
.$$
 In both cases, $m_{\mu}(V(\bp)) =m_{\mu}(V(\bq))$, so 1.8(i) is satisfied.

We have now shown that, if $V(\bp)$ is minimal, then, for some $r\ne s$ in
$\{1,2,3\}$, $\bp\in\calp^{\lambda}_{r,s}$, where
$$\calp^{\lambda}_{r,s} =\{\bp\in\calp^{\lambda}\mid V(\bp_{I\backslash I_r}) \
 \text{and $V(\bp_{I\backslash I_s})$ are both minimal}\}.$$
Note that, by 1.7, 3.1 and 5.1, if $\bp ,\bq \in\calp^{\lambda}_{r,s}$, then
$[V(\bp)] = [V(\bq)]$. Moreover, if $\bp\in\calp^{\lambda}_{r,s}$ and
$t\in\{1,2,3\}\backslash\{r,s\}$, then, by 3.1, $V(\bp_{I\backslash I_t})$ is
not minimal, and hence  by 6.2(i),
$$\align m_{\lambda-\alpha_{i_0}-\alpha_{i_r}-\alpha_{i_s}-\theta_r(\lambda
)-\theta_s(\lambda)}(V(\bp)) & > 0,\\
m_{\lambda-\alpha_{i_0}-\alpha_{i_r}-\alpha_{i_t}-\theta_r(\lambda
)-\theta_t(\lambda)}(V(\bp)) & =0,\\
m_{\lambda-\alpha_{i_0}-\alpha_{i_t}-\alpha_{i_s}-\theta_t(\lambda
)-\theta_s(\lambda)}(V(\bp)) & = 0.\endalign $$
It follows that, if $\bp^{r,s}\in\calp^{\lambda}_{r,s}$, then the
$[V(\bp^{r,s})]$, for $r<s$ in $\{1,2,3\}$, are distinct elements of
$\calq^{\lambda}$. We prove that all three elements are minimal. For this, it
suffices to show that none of them is strictly less than the other two.

Suppose, for example, that $[V(\bp^{1,2})]\prec [V(\bp^{1,3})]$. Since
$$\align m_{\lambda-\alpha_{i_0}-\alpha_{i_1}-\alpha_{i_2}-\theta_1(\lambda
)-\theta_2(\lambda)}(V(\bp^{1,2})) & > 0,\\
m_{\lambda-\alpha_{i_0}-\alpha_{i_1}-\alpha_{i_2}-\theta_1\lambda
)-\theta_2(\lambda)}(V(\bp^{1,3})) & = 0,\endalign$$
there exists $\eta\in Q^+$ such that $\eta <
\alpha_{i_0}+\alpha_{i_1}+\alpha_{i_2}+\theta_1(\lambda)+\theta_2(\lambda)$
and $m_{\lambda-\eta}(V(\bp^{1,3}))$ $> m_{\lambda-\eta}(V(\bp^{1,2}))$.
But this is impossible, since $V(\bp^{1,3}_{I\backslash I_3})$ is minimal, so
by 2.4 and 3.3, $m_{\lambda-\eta}(V(\bp^{1,3}))=0$.

The proof of Theorem 6.1 is complete.
\qed\enddemo

\vskip36pt\noindent{\bf References}
\vskip12pt\noindent
1. Beck, J., Braid group action and quantum affine algebras, preprint, MIT,
1993.

\noindent 2. Chari, V., Minimal affinizations of representations of quantum
groups: the rank 2 case, preprint, 1994.

\noindent 3.  Chari, V. and Pressley, A. N., Quantum affine algebras, Commun.
Math. Phys. {\bf 142} (1991), 261-83.

\noindent 4.  Chari, V. and Pressley, A. N., Small representations of quantum
affine algebras, Lett. Math. Phys. {\bf 30} (1994), 131-45.

\noindent 5.  Chari, V. and Pressley, A. N., {\it A Guide to Quantum Groups},
Cambridge University Press, Cambridge, 1994.

\noindent 6.  Chari, V. and Pressley, A. N., Quantum affine algebras and their
representations, preprint, 1994.

\noindent 7. Drinfel'd, V. G., A new realization of Yangians and quantized
affine algebras, Soviet Math. Dokl. {\bf 36} (1988), 212-6.

\noindent 8. Lusztig, G., {\it Introduction to Quantum Groups}, Progress in
Mathematics 110, Birkh\"auser, Boston, 1993.

\enddocument